# Origin and hole-density dependence of the intrinsic spin Hall conductivity in inorganic halide perovskites

*Thanh Tung Huynh[1,*], Yuefeng Yin[1,*], Weiyao Zhao[1], Junlin Yan[1], Qurat Ul Ain[1], Jacek J. Jasieniak[1], Julie Karel[1] and Nikhil V. Medhekar[1,2,*]*

[1] Department of Materials Science and Engineering, Monash University, Clayton, Victoria 3800, Australia

[2] School of Physics and Astronomy, Monash University, Clayton, Victoria 3800, Australia
* Corresponding authors. E-mail addresses: tung.huynh1@monash.edu (T.T. Huynh), yuefeng.yin@monash.edu (Y. Yin), nikhil.medhekar@monash.edu (N.V. Medhekar).

**Funding:** This work was supported by the Australian Research Council [grant number DP220103783]

## Abstract

Metal halide perovskites have spin–orbit coupling among the strongest in semiconductors, yet whether their centrosymmetric bulk phases can convert charge current into spin current, and what controls the response, is unknown. Here we identify, from first-principles calculations on the four all-inorganic compounds $CsBX_3$ (B = Pb, Sn; X = Br, I), the factors that govern their intrinsic spin Hall conductivity. The conductivity arises from the spin Berry curvature of the spin–orbit-split halide-*p* bands, 1.1–2.6 eV below the valence band maximum, and exceeds that of III-V and group-IV semiconductors when the Fermi level lies inside them. Because the valence band maximum is an *s*–*p* singlet, shifted only in second order, these contributions cancel in the undoped compounds, which respond like conventional semiconductors. With hole doping the response recovers, and at densities reachable by electric double-layer gating $CsPbI_3$ attains the spin Hall conductivity of *p*-GaAs. The response is therefore determined by the position of the halide-*p* bands relative to the valence band maximum, not by the atomic spin–orbit constant, a criterion that applies to any semiconductor with an orbital-singlet band edge. These results point to $CsPbI_3$ as a candidate gate-controlled source of spin current for perovskite spin–orbit torque devices.

## 1. Introduction

Spin–orbit coupling in metal halide perovskites (MHPs) $ABX_3$ is among the strongest in any semiconductor family. The atomic spin–orbit constant $\zeta$ of the B-site and halide $p$ orbitals ranges from ~0.3 eV for Sn and Br to ~1.2 eV for Pb, compared with 0.2 eV or less for the band-edge orbitals of GaAs and Si (Table 1). This coupling lowers the conduction band minimum by up to 1 eV in MHPs, and determines the band gap of the solar cells, light-emitting diodes and photodetectors made from them.[1–3] Spintronic applications rely on the same coupling to convert a charge current into a spin current through the spin Hall effect, where the spin current switches the magnetic layer in a spin–orbit torque device. In group-IV and III-V semiconductors the intrinsic spin Hall conductivity grows with the spin–orbit splitting of the valence band [4], and among the elemental metals the largest values belong to the heaviest, with platinum the best known.[5] By this measure MHPs should be among the strongest intrinsic spin Hall materials, but whether they are is not decided by $\zeta$ alone.

Research to date on spin–orbit effects in MHPs has largely focused on the Rashba effect [6–8], in which spin–orbit coupling lifts the spin degeneracy of the bands in a crystal without inversion symmetry. The bulk phases of the Pb and Sn compounds (cubic, tetragonal and orthorhombic) are centrosymmetric. In the three-dimensional hybrid perovskite $MAPbI_3$, accordingly, the static and dynamic Rashba splittings are small.[9] Large splittings have been reported in two-dimensional hybrid perovskites, where the layer structure breaks inversion symmetry.[7, 8] All-inorganic MHPs (for example, $CsPbX_3$ and $CsSnX_3$) have neither an organic cation nor a layer structure to break the symmetry, and bulk Rashba splitting is forbidden.[10–13] The spin Hall effect, by contrast, does not require broken inversion symmetry. Its intrinsic contribution arises from the spin Berry curvature of the occupied bands, which any crystal with spin–orbit coupling possesses whether or not it has inversion symmetry. In a centrosymmetric MHP, spin–orbit coupling therefore enters spin transport through the spin Hall effect, which is quantified by the intrinsic spin Hall conductivity.

The intrinsic spin Hall conductivity has been calculated and measured for a range of metals, semiconductors and oxide perovskites. The reported values, quoted here in $(\hbar/e)$ S/cm, provide a benchmark for MHPs. In platinum, the reference metal for the spin Hall effect, it is ~2000 at low temperature and ~200 at room temperature.[5] In the hole-doped semiconductors Ge, GaAs and AlAs it is ~100 [4], and in undoped Si ~3.5.[14] Among the oxide perovskites, a tight-binding model of the semimetal $SrIrO_3$ predicts ~100–130 at the Fermi level [15], and measurements on $SrRuO_3$/Co bilayers report ~200 at room temperature and more than ~1500 at 60 K.[16] For MHPs, the recent study by Abdelwahab et al. reported a spin Hall angle of 5% at room temperature in the chiral two-dimensional lead iodide perovskite $(R/S\text{-}MB)_2(MA)_3Pb_4I_{13}$, about thirty times that of its racemic counterpart. They attribute the difference to the inversion-symmetry breaking by the chiral cation.[17] To our knowledge, the bulk, three-dimensional compounds that dominate the device literature have not been examined in this respect. Whether these centrosymmetric compounds host an intrinsic spin Hall effect at realistic carrier densities, and which features of the electronic structure control its magnitude, is yet to be established.

In this work, we address both questions with first-principles calculations of the intrinsic spin Hall conductivity in the four all-inorganic compounds $CsBX_3$ (B = Pb, Sn; X = Br, I) in the orthorhombic *Pnma* structure, using density functional theory with spin–orbit coupling and maximally localised Wannier functions. The four compounds differ only in the B-site metal and the halide, which allows the dependence of the response on each to be isolated. Biaxial strain applied to $CsSnI_3$ determines the sensitivity of the response to the lattice distortion of an epitaxial film. We report the conductivity of the undoped compounds, its dependence on hole density up to the filling at which it reaches its maximum, and the separate contributions of the upper valence band and the deeper halide-*p* states. We find that the spin Berry curvature of the occupied bands is larger than in the III-V semiconductors, and that the conductivity at a given hole density is governed by the position of the spin–orbit-split states relative to the valence band edge rather than by the spin–orbit constant. In $CsBX_3$ the valence band edge is a B-*s*/halide-*p* singlet that spin–orbit coupling does not split, and the split halide-*p* states begin ~0.4–2 eV below it, whereas in GaAs the spin–orbit-split states lie at the valence band edge. Biaxial strain of the magnitude typically found in epitaxially grown thin films changes the response at accessible hole densities by at most about a factor of two. The position of the spin–orbit-split states relative to the band edge, rather than the atomic number of the constituent elements, is therefore the key property to be engineered in a halide perovskite for charge-to-spin conversion.

## 2. Results

### *2.1. Band structure and the effect of spin–orbit coupling on the band edges*

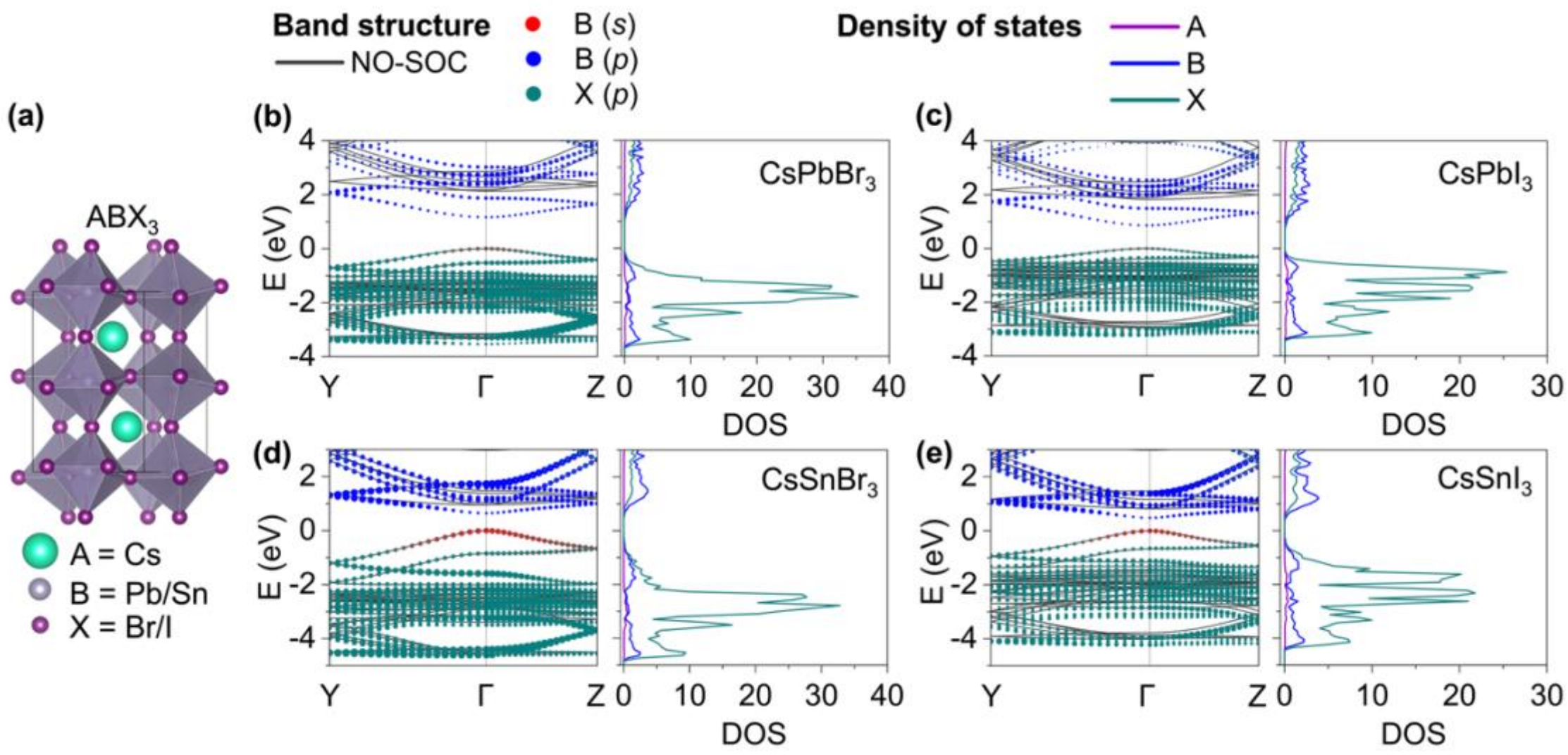

***Figure 1. Crystal structure and electronic bands of the four $CsBX_3$ compounds with and without spin–orbit coupling.*** *(a) Orthorhombic Pnma unit cell of $CsBX_3$ with the $BX_6$ octahedra shaded. (b–e) Band structures of $CsPbBr_3$, $CsPbI_3$, $CsSnBr_3$ and $CsSnI_3$ along Y–Γ–Z without spin–orbit coupling (grey lines) and with spin–orbit coupling (symbols sized by orbital weight on the same scale for all orbitals), and the corresponding element-projected densities of states with spin–orbit coupling. The valence band maximum is set to zero.*

Since the intrinsic spin Hall conductivity is a property of the occupied bands, we first establish the electronic band structure of the four MHP compounds and the effect of spin–orbit coupling on each band edge. Figure 1b–e shows the band structures of the four compounds with and without spin–orbit coupling. In every compound the valence band maximum is composed primarily of halide *p* orbitals (Br 4*p* or I 5*p*), with a small contribution from the B-site *s* orbitals (Pb 6*s* or Sn 5*s*). The conduction band minimum is dominated by B-site *p* orbitals, Pb 6*p* in the Pb compounds and Sn 5*p* in the Sn compounds. The halide-*p* bands form a dense set of states beginning ~0.4–2 eV below the valence band maximum, shallowest in $CsPbI_3$ and deepest in $CsSnBr_3$ (Figure 1b–e). When spin–orbit coupling is included, the valence band maximum remains almost unchanged whereas the conduction band minimum is lowered: the gap falls from 2.13 to 1.16 eV in $CsPbBr_3$, 1.83 to 0.86 eV in $CsPbI_3$, 0.95 to 0.65 eV in $CsSnBr_3$ and 0.82 to 0.48 eV in $CsSnI_3$ (Table S1). The shift of the conduction band minimum is ~1 eV in both Pb compounds and ~0.3 eV in both Sn compounds, and does not depend on the halide. These band gaps and their reduction by spin–orbit coupling are consistent with earlier first-principles results for the four compounds (Table S1). The calculated gaps are below the measured values, as expected for the PBE functional. The effect of the gap on the spin Hall conductivity is small and is assessed in Section 2.4.

The magnitude of the shift of the conduction band minimum induced by spin–orbit coupling, and the absence of a corresponding shift of the valence band maximum, are both determined by the orbital degeneracy of the two band-edge states rather than by the atomic number of the heaviest element. The conduction band minimum is the B-site *p* triplet ($t_{1u}$ at the R point of the cubic cell, folded to Γ in *Pnma*). Under $H_{SO} = \zeta\, L \cdot S$ a *p* triplet splits in first order, the $j = 1/2$ branch falling by $\zeta$ and the $j = 3/2$ branch rising by $\zeta/2$, so the conduction band minimum drops by $\zeta_B$ times the B-*p* weight of the state. The computed shifts of the conduction band minimum are 84–97% of the atomic spin–orbit constants of Pb 6*p* and Sn 5*p* (Table 1), as expected for a state whose B-*p* weight is less than one, and are independent of the halide because the conduction band minimum has no halide-*p* character. In contrast, the valence band maximum is an orbital singlet ($a_{1g}$ at R) with quenched angular momentum, and spin–orbit coupling acts on it only in second order through the halide-*p* bands below. It therefore moves little, even in $CsSnI_3$, where $\zeta$ of I 5*p* (0.63 eV) is almost twice that of Sn 5*p*. The role of orbital degeneracy identified here for the Pb compounds agrees with the analysis of Even et al. [18], and our four-compound comparison in Figure 1 establishes that it holds quantitatively for the Sn compounds as well.

Two features of the band structures presented in Figure 1 determine the spin Hall response discussed in the following sections. First, the spin–orbit-split states of the valence band are the halide-*p* states, which begin ~0.4–2 eV below the valence band maximum. In GaAs, by contrast, the valence band maximum is itself the *p* triplet, so the spin–orbit-split states lie at the band edge [19], and the two families differ in the orbital character of the band edge rather than in the strength of the spin–orbit coupling. Second, with spin–orbit coupling, the valence band maximum and the conduction band minimum are separated by 0.48–1.16 eV at Γ. This pair of states, an *s*–*p* singlet and a spin–orbit-split *p* state, generates the largest spin Berry curvature in the valence band. As shown below, the spin Hall conductivity at a given hole density follows the first feature rather than the second.

***Table 1. Atomic spin–orbit constants (ζ) of the band-edge orbitals and the spin–orbit shift of the conduction band minimum.*** *ζ is obtained from the spectroscopic fine-structure interval Δ of the atom or ion as $\zeta = 2\Delta/3$ for a single p electron or hole (Table S3). For the MHPs the last column is the shift of the conduction band minimum, the difference between the band gaps with and without spin–orbit coupling (Table S1). For GaAs and Si, it is the splitting of the valence band maximum, where the degenerate p state lies in those semiconductors.* [19, 20]

| Orbital | $\zeta$ (eV) | Compound | Spin–orbit shift or splitting of the $p$ band edge (eV) |
|---|---|---|---|
| Pb 6$p$ | 1.16 | $CsPbBr_3$ | 0.97 |
| | | $CsPbI_3$ | 0.97 |
| Sn 5$p$ | 0.35 | $CsSnBr_3$ | 0.30 |
| | | $CsSnI_3$ | 0.34 |
| I 5$p$ | 0.63 | | – |
| Br 4$p$ | 0.30 | | – |
| As 4$p$ | 0.21 | GaAs | 0.34 |
| Ga 4$p$ | 0.07 | GaAs | – |
| Si 3$p$ | 0.024 | Si | 0.043 |

### *2.2. Intrinsic spin Hall conductivity*

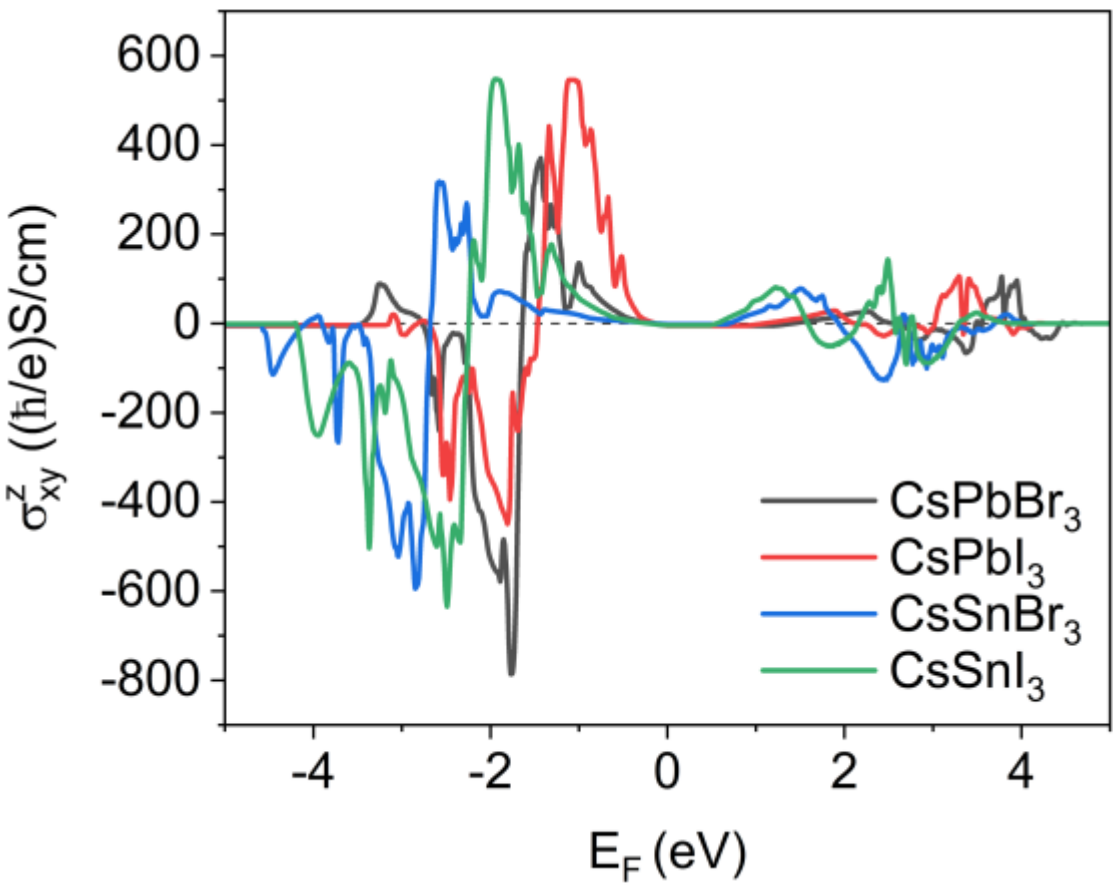

***Figure 2. Intrinsic spin Hall conductivity of the four $CsBX_3$ compounds as a function of the Fermi level.*** *The component $\sigma^z_{xy}$ of $CsPbBr_3$, $CsPbI_3$, $CsSnBr_3$ and $CsSnI_3$ from Equation 1, with the Fermi level varied through the fixed band structure. The valence band maximum is set to zero.*

Figure 2 shows the calculated spin Hall conductivity $\sigma^z_{xy}$ of the four MHP compounds as a function of the Fermi level. With the Fermi level in the gap, which corresponds to the undoped compound, $\sigma^z_{xy}$ is between −2 and −5 ($\hbar/e$) S/cm for the four compounds. The corresponding values for undoped Si and GaAs, calculated from the same Kubo formula, are ~3.5 and ~22 ($\hbar/e$) S/cm [14], so the in-gap response of the four MHP compounds is comparable in magnitude to that of conventional semiconductors. Yao and Fang [14] attribute this in-gap value to a spin Hall insulator response that produces no spin accumulation, and subtract it as a baseline before comparing with experiment. Since its magnitude is at most 5 ($\hbar/e$) S/cm for MHPs, whether or not it is subtracted does not affect the comparison that follows. The maximum values of $\sigma^z_{xy}$ are 370, 546, 318 and 548 ($\hbar/e$) S/cm for $CsPbBr_3$, $CsPbI_3$, $CsSnBr_3$ and $CsSnI_3$ (Table S2). These are two to four times the maximum calculated values for III-V and group-IV semiconductors, ~130 for GaAs and ~100 for Ge and AlAs.[14, 4] Importantly, the two families reach their maxima at different positions of the Fermi level. In GaAs the maximum is reached with the Fermi level ~0.15 eV below the valence band maximum [14], at a hole density of

~8×10[20] cm[−3].[4] In the four MHP compounds it is reached 1.1–2.6 eV below the valence band maximum, at hole densities of 2.4–3.6×10[22] cm[-3] (Table S4), which corresponds to ~6–7 holes per formula unit. Such densities are beyond the reach of doping or gating, so the maxima measure the capacity of the band structure rather than an accessible response. The accessible range, up to ~3×10[21] cm[−3], is examined in Table 2.

***Table 2. Intrinsic spin Hall conductivity and Fermi level of the four $CsBX_3$ compounds at fixed hole density.*** *For each compound the first column is $\sigma^z_{xy}$ in (ħ/e) S/cm and the second is the position of the Fermi level below the valence band maximum, $E_F$ (eV), obtained by integrating the density of states from the valence band maximum. The values for hole-doped GaAs are from Guo et al.* [4]

| Hole density ($cm^{-3}$) | $CsPbBr_3$ | | $CsPbI_3$ | | $CsSnBr_3$ | | $CsSnI_3$ | | *p*-GaAs |
|---|---|---|---|---|---|---|---|---|---|
| | $\sigma^z_{xy}$ | $E_F$ | $\sigma^z_{xy}$ | $E_F$ | $\sigma^z_{xy}$ | $E_F$ | $\sigma^z_{xy}$ | $E_F$ | $\sigma^z_{xy}$ |
| $10^{19}$ | −4.0 | 0.04 | −3.6 | 0.04 | −0.1 | 0.12 | −1.8 | 0.12 | – |
| $10^{20}$ | 0.1 | 0.19 | 1.8 | 0.16 | 3.7 | 0.37 | 4.2 | 0.36 | ~70 |
| $10^{21}$ | 20.0 | 0.52 | 29.4 | 0.36 | 12.5 | 0.88 | 25.8 | 0.74 | ~130 |
| $3×10^{21}$ | 54.2 | 0.76 | 110.5 | 0.49 | 27.2 | 1.27 | 63.6 | 1.03 | ~110 |

Table 2 presents the spin Hall conductivity $\sigma^z_{xy}$ at fixed hole density, with the Fermi level at each density obtained from the integrated density of states of each MHP compound. At a hole density of $10^{19}$ cm$^{-3}$ the Fermi level lies within 0.12 eV of the valence band maximum and $\sigma^z_{xy}$ is within ~2 (*ħ*/*e*) S/cm of its in-gap value in all four compounds. At $10^{20}$ cm$^{-3}$ it is 0–4 (*ħ*/*e*) S/cm, and at $10^{21}$ cm$^{-3}$ it rises to 12–29 (*ħ*/*e*) S/cm, with $CsPbI_3$ the highest of the four. Over the same range of hole densities, *p*-GaAs rises from ~70 to ~130 (*ħ*/*e*) S/cm.[4] At $3×10^{21}$ cm$^{-3}$, a density that an electric double layer created by ionic liquid gating can reach in the first few nanometres of a film (sheet densities of $10^{14}$–$10^{15}$ cm$^{-2}$ [21]), $\sigma^z_{xy}$ is 27–110 (*ħ*/*e*) S/cm, and the value for $CsPbI_3$ is the same as that of *p*-GaAs at the same density, ~110 (*ħ*/*e*) S/cm (Table 2).[4]

Two further features of Figure 2 are used in the analysis below. First, the maximum values of $\sigma^z_{xy}$ are ordered by the halide. The two iodides reach 548 and 546 (*ħ*/*e*) S/cm and the two bromides 370 and 318 (*ħ*/*e*) S/cm, whereas at fixed halide the Pb and Sn compounds differ by at most ~15%. In each compound the maximum lies within the halide-*p* bands identified in Figure 1, 1.1–2.6 eV below the valence band maximum, rather than at the band edge. Second, the three calculated components $\sigma^z_{xy}$, $\sigma^x_{yz}$ and $\sigma^y_{zx}$ show a similar dependence on the Fermi level in every compound (Figure S4), with maximum values that differ by 8–26% (Table S2). We report here only the values of $\sigma^z_{xy}$, which has the smallest maximum of the three. In the next section, we examine why $\sigma^z_{xy}$, which reaches 318–548 (*ħ*/*e*) S/cm with the Fermi level inside the valence band, falls to a few (*ħ*/*e*) S/cm when the valence band is fully occupied.

### *2.3. Origin of the spin Hall conductivity*

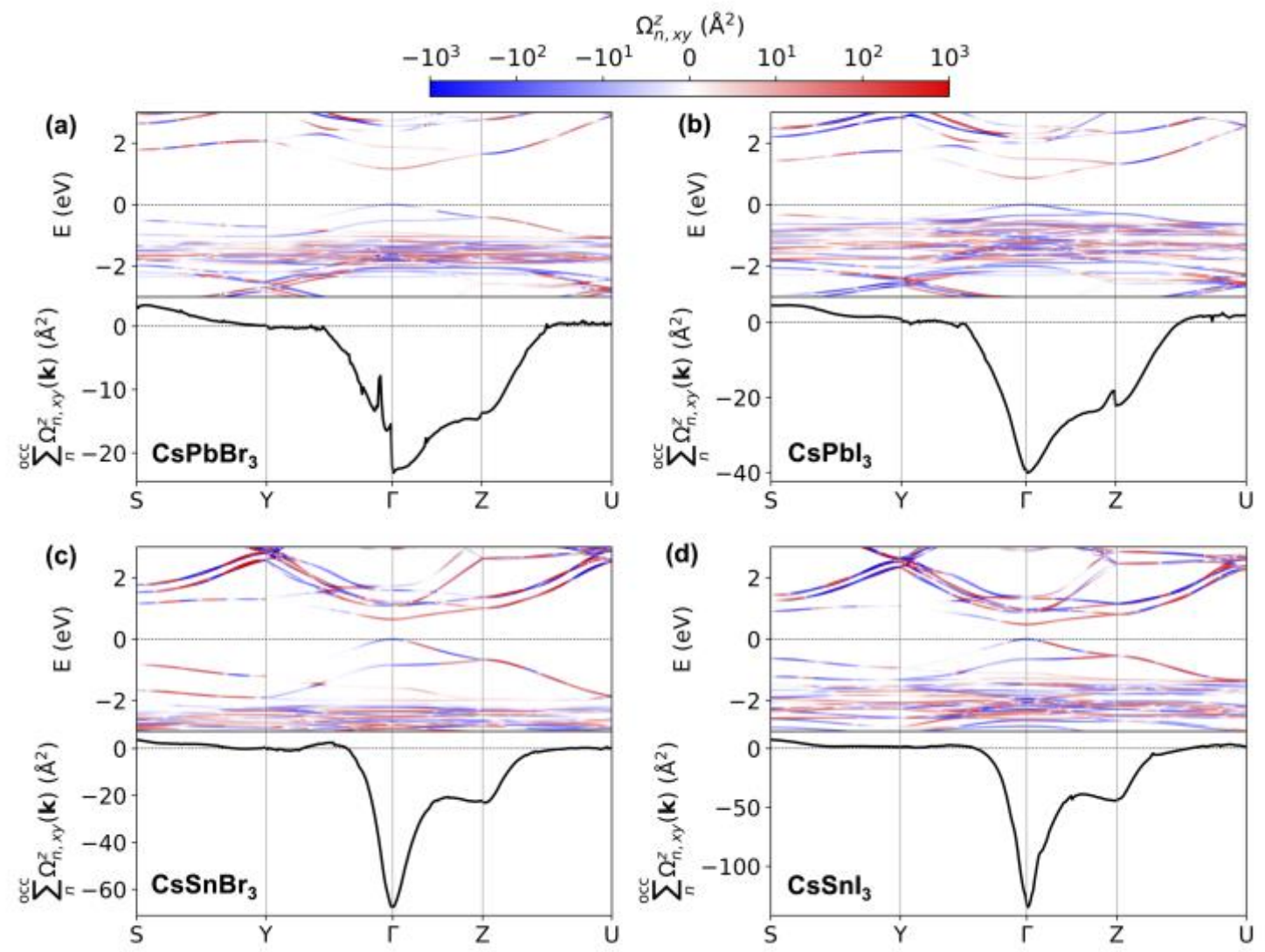

***Figure 3. Spin Berry curvature of the four $CsBX_3$ compounds.*** *(a–d) Band structures of $CsPbBr_3$, $CsPbI_3$, $CsSnBr_3$ and $CsSnI_3$ along S–Y–Γ–Z–U with spin–orbit coupling, each state coloured by its spin Berry curvature $\Omega^z_{n,xy}(\mathbf{k})$ on a logarithmic scale for each sign (upper panels), and $\Omega^z_{xy}(\mathbf{k})$ summed over the valence bands (lower panels). The valence band maximum is set to zero.*

The dependence of the spin Hall conductivity $\sigma^z_{xy}$ on the Fermi level in Figure 2 follows from the structure of Equation 1 and 2. In Equation 2 the sum over $m$ runs over all other bands, occupied or empty, and is independent of the Fermi level. The spin Berry curvature of a band is fixed, and the Fermi level enters only through the occupation $f_{n\mathbf{k}}$ in Equation 1. The conductivity $\sigma^z_{xy}(E_F)$ is thus a cumulative sum over the occupied states, and the change in $\sigma^z_{xy}$ between two Fermi levels is the contribution of the states between them. The states between the valence band maximum and $E_{max}$, the position of the maximum of $\sigma^z_{xy}$, contribute negatively, since removing them increases $\sigma^z_{xy}$ from its in-gap value to 318–548 ($\hbar/e$) S/cm. The states immediately below $E_{max}$ contribute positively, since removing them decreases $\sigma^z_{xy}$ to between ~−450 and ~−790 ($\hbar/e$) S/cm. Equation 2 also shows which pairs of bands contribute. A pair of bands $n$ and $m$ contributes equal and opposite amounts to $\Omega_n$ and $\Omega_m$, so when both bands are occupied their contributions cancel in Equation 1. For the undoped compound $\sigma^z_{xy}$ is therefore a sum over pairs of one occupied and one empty band only. The same cancellation holds at each $k$. In the spin Berry curvature summed over the valence bands (Figure 3, lower panels), the pair terms between two valence bands cancel, and only the pair terms between a valence band and a conduction band remain. This curvature is negative from Γ to Z in every compound, with its largest magnitude at Γ: −22, −39, −68 and −133 Å$^2$ for $CsPbBr_3$, $CsPbI_3$, $CsSnBr_3$ and $CsSnI_3$. Elsewhere along the path it is small.

The in-gap value and the maximum of $\sigma^z_{xy}$ differ by two orders of magnitude, and the dependence of the maximum on the smearing width $\eta$ identifies the pairs of bands responsible for it. At $\omega = 0$ each pair term in Equation 2 is reduced by the factor $\Delta^2/(\Delta^2 + \eta^2)$, where $\Delta$ is the separation of the two bands. In $CsSnI_3$ the maximum of $\sigma^z_{xy}$ falls from 548 ($\hbar/e$) S/cm at $\eta = 0.05$ eV to 214 at $\eta = 1$ eV (Figure S3). With the Fermi level at the maximum, the conduction bands lie more than 2.4 eV above the occupied states, and their pair terms change by less than 15% over this range of $\eta$. The dependence on $\eta$ therefore attributes more than half of the maximum to pairs of valence bands separated by less than ~1 eV, one above the Fermi level and one below it. The position of the maximum is consistent with this: in each compound $E_{max}$ lies ~0.4–0.7 eV below the top of the halide-*p* bands identified in Figure 1, which begin ~0.4 eV below the valence band maximum in $CsPbI_3$ and ~2 eV below it in $CsSnBr_3$. When every valence band is occupied, these pairs cancel, as shown above, and the in-gap value is the sum over the valence–conduction pairs alone. The spin Hall conductivity of the undoped compounds is small not because the spin Berry curvature of the valence band is small but because the pairs of bands that generate it are all occupied. The same mechanism operates in *p*-GaAs, where the spin–orbit-split bands lie at the valence band maximum and the maximum of $\sigma^z_{xy}$ is reached ~0.15 eV below it.[14, 4] In the MHP compounds the split states are the halide-*p* bands below the singlet valence band maximum, and the maximum of $\sigma^z_{xy}$ is reached 1.1–2.6 eV below the edge.

The size of the maximum and its position are governed by different constituents of the compound. The size is ordered by the halide, as noted above (Figure 2), which indicates that the pair terms within the halide-*p* bands arise from the spin–orbit coupling of the halide. Guo et al. found the same dependence across Si, Ge, GaAs and AlAs, where the intrinsic spin Hall conductivity grows with the spin–orbit splitting of the valence band.[4] The position is fixed by the depth at which the halide-*p* bands begin below the valence band maximum, taken as the top of the dense set of flat halide-*p* bands in Figure 1. This depth is ~0.4 and ~1.0 eV in $CsPbI_3$ and $CsPbBr_3$ and ~1.4 and ~2.0 eV in $CsSnI_3$ and $CsSnBr_3$, and so depends on the B site. $CsPbI_3$ combines the larger spin–orbit coupling of I 5*p* with the shallowest halide-*p* bands, and has the largest $\sigma^z_{xy}$ of the four at hole densities of $10^{21}$ cm$^{-3}$ and above (Table 2). The spin Berry curvature at Γ in Figure 3, which arises from the pair terms across the gap, scales as $1/\Delta^2$ with the band gap: $|\Omega_\Gamma|\Delta^2 = 29$–$31$ Å$^2$ eV$^2$ in the four compounds. However, the contribution of these pair terms to $\sigma^z_{xy}$ is small: neither the in-gap values nor the values at fixed hole density in Table 2 follow the gap, which varies by a factor of 2.4 across the four compounds. The spin Hall conductivity of the MHP compounds at a given hole density is therefore controlled by the position of the halide-*p* bands relative to the valence band maximum rather than by the band gap at Γ. In the next section, we test this conclusion with biaxial strain, which shifts the halide-*p* bands.

### *2.4. Spin Hall conductivity under biaxial strain*

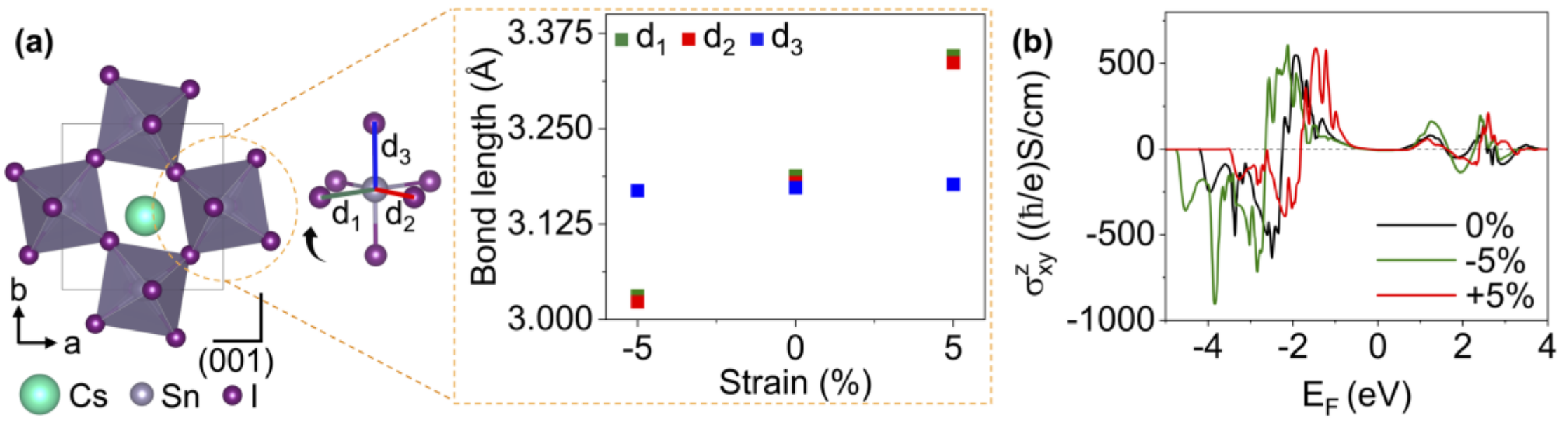

***Figure 4. Effect of biaxial strain on the structure and spin Hall conductivity of $CsSnI_3$.*** *(a) Sn–I bond lengths of $CsSnI_3$ under biaxial strain in the (001) plane, with the unit cell viewed along c and the three inequivalent bonds $d_1$ and $d_2$ (in-plane) and $d_3$ (out-of-plane) marked on the $SnI_6$ octahedron. (b) $\sigma^z_{xy}$ as a function of the Fermi level at 0, −5 and +5% strain. The valence band maximum is set to zero.*

Thin films of MHPs on a substrate can be subject to biaxial strain, from the lattice mismatch in epitaxial films [22] and from substrate clamping on cooling.[23] We therefore investigate the effect of biaxial strain on the spin Hall conductivity, with $CsSnI_3$ as a representative compound. Figure 4b shows $\sigma^z_{xy}$ of $CsSnI_3$ at 0, −5 and +5% biaxial strain in the (001) plane, and Figure S5 shows the same for ±3% and ±7% and for the (010) and (100) planes. Under compressive strain $E_{max}$ moves deeper below the valence band maximum, by ~0.2 eV at −5%, and under tensile strain it moves towards the valence band maximum, by ~0.5 eV at +5%. In the (001) plane the maximum value of $\sigma^z_{xy}$ increases by up to ~10% at ±5% (Figure 4b). In the (010) and (100) planes it increases under compressive strain and decreases under tensile strain, by up to ~15% at ±7% (Table S6). With the Fermi level in the gap, $\sigma^z_{xy}$ remains within a few ($\hbar/e$) S/cm of its unstrained value at every strain in the three planes (Table S6). Near the valence band maximum the response changes little for strains within ±3%, the range imposed by lattice mismatch in epitaxial films.[22] At the Fermi level that corresponds to $10^{21}$ $cm^{-3}$ in the unstrained compound, 0.74 eV below the valence band maximum, it is unchanged in the (001) plane. In the (010) and (100) planes it changes by less than a factor of two, and it decreases under compressive strain and increases under tensile strain (Figure S5, Table S6). Larger tensile strain increases $\sigma^z_{xy}$ at every hole density above ~$10^{21}$ $cm^{-3}$, since it brings $E_{max}$ towards the valence band maximum. At +5% in the (001) plane, $\sigma^z_{xy}$ at 1.0 eV below the valence band maximum (a hole density of $2.7\times10^{21}$ $cm^{-3}$ in the unstrained compound) is more than three times its unstrained value. Compressive strain of 5% lowers it by 10–50% depending on the plane.

Under strain, the in-plane Sn–I bonds follow the lattice, from 3.19 Å unstrained to 3.03 Å at −5% and 3.34 Å at +5%, whereas the out-of-plane bond is unchanged at 3.17 Å (Figure 4a). The Sn–I–Sn angle changes from 161.5° to between 152.6° and 168.3° at ±5% in the three planes (Table S5). The change in $\sigma^z_{xy}$ with strain follows from the origin of the two parts of the response identified in the previous section. The maximum arises from pairs within the halide-*p* bands, which strain shifts as a whole. Compressive strain moves these bands and $E_{max}$ with them deeper below the valence band maximum, and tensile strain moves them towards it. The in-gap value arises from the pair terms between the valence and conduction bands and is unchanged, even though the spin Berry curvature at Γ changes from −133 Å² unstrained to ~−175 Å² at −5% and ~−63 Å² at +5% with the change in the gap at Γ, from 0.48 eV unstrained to 0.29 eV at −5% and 0.63 eV at +5% (Figure S6). This confirms that the pair terms across the gap contribute little to $\sigma^z_{xy}$. The response near the valence band maximum follows the halide-*p* bands: when tensile strain brings them towards the valence band maximum, the pair terms between the top valence band and the halide-*p* bands below it grow, and $\sigma^z_{xy}$ at a given hole density rises. Since the same two parts make up the response in all four MHP compounds, we expect the same behaviour in each. Tensile strain therefore acts on $\sigma^z_{xy}$ at a given hole density in the same way as the choice of compound in the previous section, through the position of the halide-*p* bands relative to the valence band maximum.

It is worth pointing out the approximations on which these results rest. Our calculations give the intrinsic spin Hall conductivity only. The extrinsic contributions from skew scattering and side jump at impurities, which depend on the sample [24], are not included. The dependence on hole density is obtained in the rigid-band approximation, with the Fermi level moved through the bands of the undoped compound, as in the calculations for the III-V and group-IV semiconductors with which we compare.[4, 14] The PBE band gap is below the measured gap, 0.48 eV compared with 1.3 eV for $CsSnI_3$, and HSE06 gives 0.86 eV (Methods). The pair terms across the gap scale as $1/\Delta^2$, so at the HSE06 gap they would be reduced to ~30% of their PBE value. Their contribution to $\sigma^z_{xy}$ is small: under 5% tensile strain the spin Berry curvature at Γ halves, from −133 to ~−63 Å², while the in-gap value changes by ~1 ($\hbar/e$) S/cm (Table S6,

Figure S6), so at the HSE06 gap the change would be a few ($\hbar/e$) S/cm. The in-gap values and the entries of Table 2 at $10^{19}$ and $10^{20}$ cm$^{-3}$, which are themselves a few ($\hbar/e$) S/cm, are therefore uncertain in sign. The entries at $10^{21}$ cm$^{-3}$ and above and the maxima, which arise from pairs within the valence band, are not affected. The PBE volumes are 4–6% above the measured volumes (Table S1), a lattice change of ~2%. Table S6 shows that a 3% change moves $E_{\mathrm{max}}$ by up to 0.3 eV and changes $\sigma^{z}_{xy}$ at $10^{21}$ cm$^{-3}$ by less than a factor of two. The structures considered here are static and at 0 K. The thermal motion of the octahedra and the dynamic off-centering of $Sn^{2+}$ reported for $CsSnBr_3$ above 290 K [12] are expected to shift the band edges. Notwithstanding these limitations, the mechanism identified here and the ordering of the four compounds follow from the structure of Equation 2 and from the band structures of Figure 1.

## 3. Conclusions

In summary, we have calculated the intrinsic spin Hall conductivity of the four orthorhombic MHP compounds $CsBX_3$ (B = Pb, Sn; X = Br, I) from first principles and traced it to the spin Berry curvature of the valence bands. We find that the maximum spin Hall conductivity of the four compounds, 318–548 ($\hbar/e$) S/cm, is two to four times the values calculated for III-V and group-IV semiconductors. The maximum conductivity is reached with the Fermi level 1.1–2.6 eV below the valence band maximum, inside the spin–orbit-split halide-*p* bands, because the valence band maximum itself is an *s*–*p* singlet on which spin–orbit coupling acts only in second order. It arises from pairs of halide-*p* bands, one emptied and one occupied. These pairs cancel when the valence band is full, so that the spin Hall conductivity of the undoped compounds is a few ($\hbar/e$) S/cm, as in Si and GaAs. Hole doping of MHPs increases the spin Hall conductivity to 12–29 ($\hbar/e$) S/cm at $10^{21}$ cm$^{-3}$ and to 27–110 ($\hbar/e$) S/cm at $3\times10^{21}$ cm$^{-3}$, a density within reach of an electric double layer created by ionic liquid gating in the surface layer of a film [21]. At this density the spin Hall conductivity of $CsPbI_3$ is comparable to that of *p*-GaAs. Biaxial strain of up to 3%, the range reached in epitaxial films [22], changes the response at hole densities up to ~$10^{21}$ cm$^{-3}$ by less than a factor of two, and tensile strain increases it at higher densities by bringing the halide-*p* bands towards the valence band maximum.

These results identify the position of the halide-*p* bands relative to the valence band maximum, rather than the atomic spin–orbit constant or the band gap, as the parameter that controls the spin Hall conductivity of a centrosymmetric MHP at a given hole density. $CsPbI_3$, in which these bands begin ~0.4 eV below the valence band maximum, is therefore the most promising of the four compounds. Tensile strain, which in $CsSnI_3$ brings the halide-*p* bands towards the valence band maximum, is expected to increase its response further. The same argument follows from Equation 1 and 2 for any semiconductor whose valence band maximum is an orbital singlet: the spin Hall conductivity at a given hole density should be governed by how far below the band edge the spin–orbit-split states lie. The dependence of the spin Hall conductivity on hole density, which persists under biaxial strain of up to 3%, permits its electrostatic control in a thin film [25, 21], a mode of control that is not available in a spin Hall metal. Together with the solution processing and the optoelectronic functions already established for MHPs [1–3], a gate-controlled spin Hall conductivity can allow a single MHP film to serve as the source of spin current for spin–orbit torque switching [26, 27] and for spin–charge interconversion [24] in perovskite optoelectronic devices.

## 4. Methods

### *4.1. Structures*

The four inorganic metal halide perovskite compounds $CsBX_3$ (B = Pb, Sn; X = Br, I) were considered in the orthorhombic *Pnma* structure (space group 62, Figure 1a), with four formula units in a unit cell. The *Pnma* phase is the lowest-temperature perovskite phase of all four compounds.[11, 12, 28, 29] It was also used for $CsSnBr_3$, which is cubic at room temperature [12], so that the four compounds differ in composition alone. The relaxed lattice parameters and volumes are compared with earlier reported theoretical predictions and experimental measurements in Table S1. Our calculated volumes are within ~6% of the measured volumes.

### *4.2. Density functional theory calculations*

First-principles density functional theory calculations were performed using the Vienna Ab initio Simulation Package (VASP) [30, 31], with the Perdew–Burke–Ernzerhof (PBE) generalised-gradient functional [32] and projector augmented-wave potentials [33]. A plane-wave cutoff of 500 eV and a Γ-centred 6×6×4 *k*-point mesh were used throughout, with the total energy converged to $10^{-4}$ eV. The lattice parameters and atomic positions were relaxed until the force on every atom was below 0.01 eV Å$^{-1}$. Spin–orbit coupling was included self-consistently in the non-collinear formalism. Since the PBE functional underestimates the band gap, we also calculated the band structure of $CsSnI_3$ with the HSE06 hybrid functional [34] to check the band dispersion. With spin–orbit coupling, HSE06 opens the gap from 0.48 to 0.86 eV, compared with the measured 1.3 eV [29], and leaves the dispersion of the bands near the gap unchanged (Figure S1). Projected band structures and densities of states were obtained with VASPKIT [35], and the structures were visualised with VESTA.[36]

### 4.3. Wannier functions

Maximally localised Wannier functions were constructed with Wannier90 [37] from initial projections generated by the selected columns of the density matrix (SCDM) method.[38, 39] The 192 lowest spinor bands of each compound were used and the number of Wannier functions was set equal to the number of bands, so that no disentanglement was needed. The spread was minimised for 200 iterations and converged. The Wannier-interpolated bands of $CsSnI_3$ reproduce the DFT bands from 3 eV below to 3 eV above the gap (Figure S2), a range that includes the energy at which the spin Hall conductivity reaches its maximum. The same agreement was obtained for the other three compounds.

### 4.4. Spin Hall conductivity

The intrinsic spin Hall conductivity was calculated from the Kubo formula in the Wannier basis as implemented in Wannier90 [40–42],

$$\sigma_{xy}^{z}(\omega) = -\frac{e^2}{\hbar}\frac{1}{VN_k}\sum_{\boldsymbol{k}}\sum_{n} f_{n\boldsymbol{k}}\,\Omega_{n,xy}^{z}(\boldsymbol{k}) \tag{1}$$

$$\Omega_{n,xy}^{z}(\boldsymbol{k}) = \hbar^2 \sum_{m\neq n} \frac{-2\,\mathrm{Im}\big[\langle n\boldsymbol{k}|\,1/2\,\{\hat{\sigma}_z,\hat{v}_x\}|m\boldsymbol{k}\rangle\langle m\boldsymbol{k}|\hat{v}_y|n\boldsymbol{k}\rangle\big]}{(\varepsilon_{n\boldsymbol{k}}-\varepsilon_{m\boldsymbol{k}})^2-(\hbar\omega+i\eta)^2} \tag{2}$$

where $V$ is the cell volume, $N_k$ is the total number of $k$-points in the interpolation mesh, $f_{n\mathbf{k}}$ is the Fermi–Dirac occupation, $½\{\hat{\sigma_z}, \hat{v}_x\}$ is the spin-current operator and $\hat{v}_y$ is the velocity operator. The band index $n$ runs over all bands and, in Equation 2, $m$ runs over all bands other than $n$, occupied or empty. The spin Hall conductivity was calculated in the static limit ($\omega$ = 0) with a fixed smearing width $\eta$ = 0.05 eV. The Brillouin-zone sum was evaluated on a 24×24×16 interpolation mesh, refined by a factor of five at $k$-points where $|\Omega|$ exceeds 100 Å$^2$.[40] The spin Hall conductivity of $CsSnI_3$ was found to be converged with respect to both parameters: it is unchanged for interpolation meshes up to 48×48×32, and its maximum changes by ~5% between $\eta$ = 0.05 and 0.1 eV (Figure S3). The calculations for Si and GaAs with which we compare used broadenings of 0–20 meV.[14] Three of the six independent components allowed by the Laue class *mmm* of *Pnma* were calculated, $\sigma^z_{xy}$, $\sigma^x_{yz}$ and $\sigma^y_{zx}$, where $\sigma^z_{xy}$ gives the spin current along $x$, with spin polarised along $z$, for an electric field along $y$.

### *4.5. Strain*

To evaluate the effect of strain on spin Hall conductivity, biaxial strain was applied to $CsSnI_3$ in the (001) plane by scaling the $a$ and $b$ lattice parameters together, with $c$ held at its relaxed value $c_0$ to isolate the effect of the in-plane strain. The strain $\varepsilon = (a - a_0)/a_0$, where $a_0$ is the relaxed lattice parameter, was set to −7, −5, −3, +3, +5 and +7%. At each strain the atomic positions were relaxed until the force on every atom was below 0.01 eV $Å^{-1}$. The same procedure was applied for strain in the (010) and (100) planes (Figure S5).

## CRediT authorship contribution statement

Thanh Tung Huynh: Investigation, Formal analysis, Visualization, Writing – original draft. Yuefeng Yin: Methodology, Supervision, Writing – review & editing. Weiyao Zhao: Writing – review & editing. Junlin Yan: Writing – review & editing. Qurat Ul Ain: Writing – review & editing. Jacek J. Jasieniak: Funding acquisition, Writing – review & editing. Julie Karel: Funding acquisition, Writing – review & editing. Nikhil V. Medhekar: Writing – review & editing, Supervision, Funding acquisition, Conceptualization.

## Declaration of generative AI and AI-assisted technologies in the manuscript preparation process

During the preparation of this work the authors used Claude (Anthropic) in order to check grammar and improve the readability of the text. After using this tool, the authors reviewed and edited the content as needed and take full responsibility for the content of the published article.

## Declaration of competing interest

The authors declare that they have no known competing financial interests or personal relationships that could have appeared to influence the work reported in this paper.

## Acknowledgements

All authors acknowledge funding support from the Australian Research Council through Discovery Project DP220103783. This research was undertaken with the assistance of resources from the National Computational Infrastructure (NCI Australia), an NCRIS enabled capability supported by the Australian Government. This work was also supported by computational resources provided by Monash University and the Pawsey Supercomputing Research Centre with funding from the Australian Government and the Government of Western Australia. T.T.H. gratefully acknowledges the Monash Research Scholarship (Materials Science & Engineering) and the Faculty of Engineering International Postgraduate Research Scholarship (FEIPRS) from Monash University.

## Data availability

Data will be made available on request.

## Supplementary Information

# Origin and hole-density dependence of the intrinsic spin Hall conductivity in inorganic halide perovskites

*Thanh Tung Huynh[1,*], Yuefeng Yin[1,*], Weiyao Zhao[1], Junlin Yan[1], Qurat Ul Ain[1], Jacek J. Jasieniak[1], Julie Karel[1] and Nikhil V. Medhekar[1,2,*]*

[1] Department of Materials Science and Engineering, Monash University, Clayton, Victoria 3800, Australia

[2] School of Physics and Astronomy, Monash University, Clayton, Victoria 3800, Australia

* Corresponding authors. E-mail addresses: tung.huynh1@monash.edu (T.T. Huynh), yuefeng.yin@monash.edu (Y. Yin), nikhil.medhekar@monash.edu (N.V. Medhekar).

***Table S1. Lattice parameters, cell volumes and band gaps of the four $CsBX_3$ compounds in the orthorhombic Pnma structure, compared with earlier calculations and measurements.*** *a, b and c are the lattice parameters and V the volume of the four-formula-unit cell. Band gaps are given without and with spin–orbit coupling (SOC). $V/V_{exp}$ is the ratio of the calculated volume to the measured volume. All lattice parameters are given in the same axis setting, with c the long axis. References [1], [3], [5], [7], [12] and (for $CsPbI_3$) [10] report b as the long axis, and their values have been permuted accordingly. Volumes are a×b×c of the lattice parameters listed. The measured cell of $CsSnBr_3$ [12] is at 100 K, and the measured gap of $CsSnBr_3$ [13] is for the cubic phase at room temperature.*

| Compound | *a* (Å) | *b* (Å) | *c* (Å) | *V* (Å³) | $V/V_{exp}$ | Gap without SOC (eV) | Gap with SOC (eV) | Source |
|---|---|---|---|---|---|---|---|---|
| **$CsPbBr_3$** | **8.2** | **8.5** | **11.9** | **829** | **1.053** | **2.13** | **1.16** | **This work** |
| | 8.5 | 8.6 | 12.2 | 892 | | 1.78 | 0.84 | DFT [1] |
| | 8.3 | 8.4 | 11.7 | 816 | | 2.65 | 1.57 | DFT [2] |
| | 8.1 | 8.3 | 11.7 | 787 | | 1.93 | 1.06 | DFT [3] |
| | – | – | – | – | | 2.11 | 1.08 | DFT [4] |
| | 8.2 | 8.5 | 11.9 | 829 | | 2.12 | – | DFT [5] |
| | – | – | – | – | | 2.34 | – | DFT [6] |
| | 8.2 | 8.2 | 11.7 | 787 | | 2.0 | – | Exp. [7] |
| **$CsPbI_3$** | **8.7** | **9.1** | **12.7** | **1005** | **1.050** | **1.83** | **0.86** | **This work** |
| | 8.7 | 8.8 | 12.5 | 957 | | 1.83 | 0.84 | DFT [8] |
| | – | – | – | – | | 1.82 | 0.78 | DFT [4] |
| | 9.0 | 8.7 | 12.6 | 987 | | 1.78 | 0.83 | DFT [9] |
| | – | – | – | – | | 1.9 | – | DFT [6] |
| | 8.7 | 9.1 | 12.6 | 998 | | 1.82 | – | DFT [10] |
| | 8.9 | 8.6 | 12.5 | 957 | | 1.81 | – | Exp. [11] |
| **$CsSnBr_3$** | **8.2** | **8.3** | **11.8** | **803** | **1.055** | **0.95** | **0.65** | **This work** |
| | – | – | – | – | | 0.93 | 0.62 | DFT [4] |
| | 8.2 | 8.4 | 11.8 | 813 | | 0.94 | – | DFT [5] |
| | 8.0 | 8.2 | 11.6 | 761 | | – | – | Exp. [12] |
| | – | – | – | – | | 1.77 | – | Exp. [13] |
| **$CsSnI_3$** | **8.7** | **8.9** | **12.5** | **968** | **1.043, 1.043** | **0.82** | **0.48** | **This work** |
| | 8.9 | 8.7 | 12.5 | 968 | | ~0.8 | ~0.5 | DFT [14] |

| Compound | *a* (Å) | *b* (Å) | *c* (Å) | *V* (Å³) | $V/V_{exp}$ | Gap without SOC (eV) | Gap with SOC (eV) | Source |
|---|---|---|---|---|---|---|---|---|
| | – | – | – | – | | 0.81 | 0.47 | DFT [4] |
| | 8.9 | 8.7 | 12.5 | 968 | | 0.81 | – | DFT [10] |
| | 8.7 | 8.6 | 12.4 | 928 | | 1.3 | – | Exp. [15] |
| | 8.7 | 8.6 | 12.4 | 928 | | 1.31 | – | Exp. [16] |

### Band structure of $CsSnI_3$ calculated using the HSE06 hybrid functional

Since the PBE functional underestimates the band gap, the band structure of $CsSnI_3$ was also calculated using the HSE06 hybrid functional [17], with and without spin–orbit coupling. HSE06 opens the gap from 0.82 to 1.22 eV without spin–orbit coupling and from 0.48 to 0.86 eV with it, compared with the measured 1.3 eV.[15] The shift of the conduction band minimum by spin–orbit coupling is 0.36 eV with HSE06 and 0.34 eV with PBE. The energies of the valence bands within 2 eV of the valence band maximum, including the onset of the halide-*p* bands at ~1.4 eV, are unchanged (Figure S1b, compared with Figure 1e of the main text). The PBE bands were therefore used for the spin Hall conductivity.

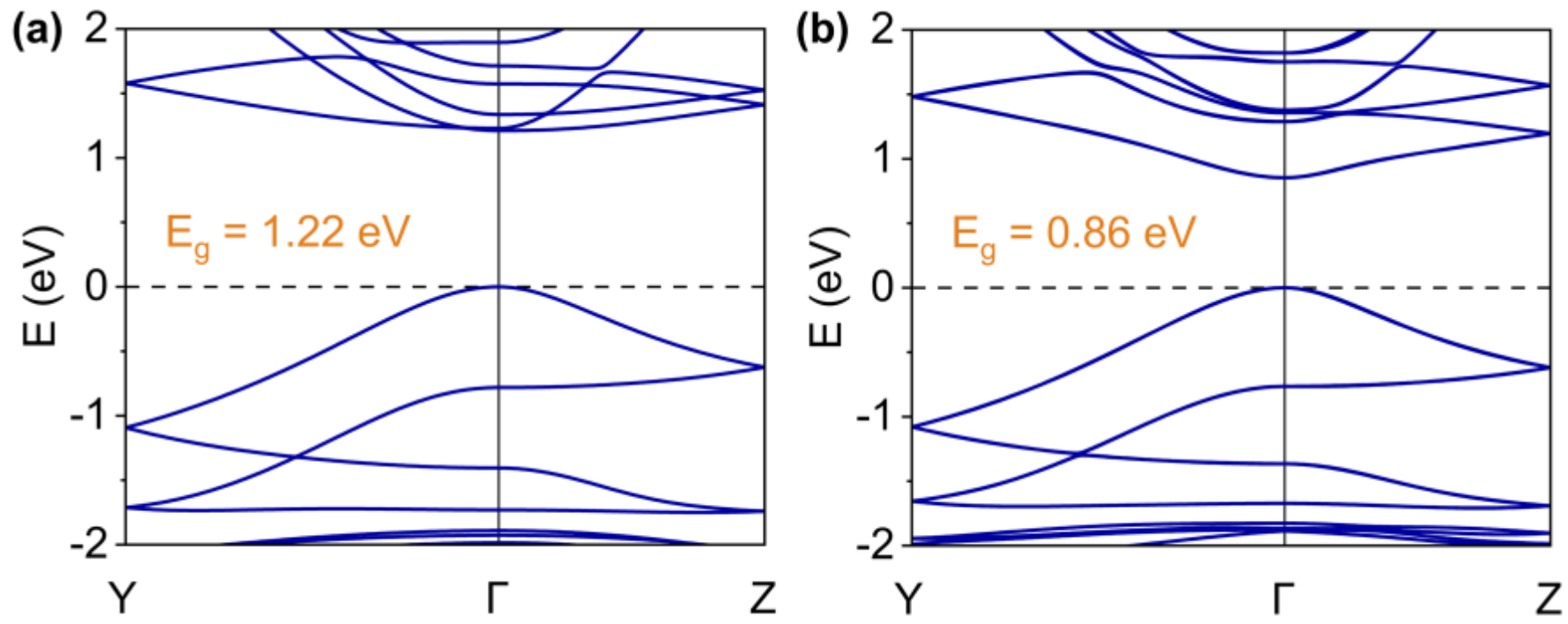

***Figure S1. Band structure of $CsSnI_3$ calculated using the HSE06 hybrid functional.*** *Bands along Y–Γ–Z (a) without and (b) with spin–orbit coupling, with the band gap* $E_g$ *marked. The valence band maximum is set to zero.*

### Wannier construction

The Wannier-interpolated bands of $CsSnI_3$ reproduce the DFT bands along the whole path from 3 eV below to 3 eV above the gap, including the halide-*p* bands (Figure S2). The same agreement was obtained for the other three compounds.

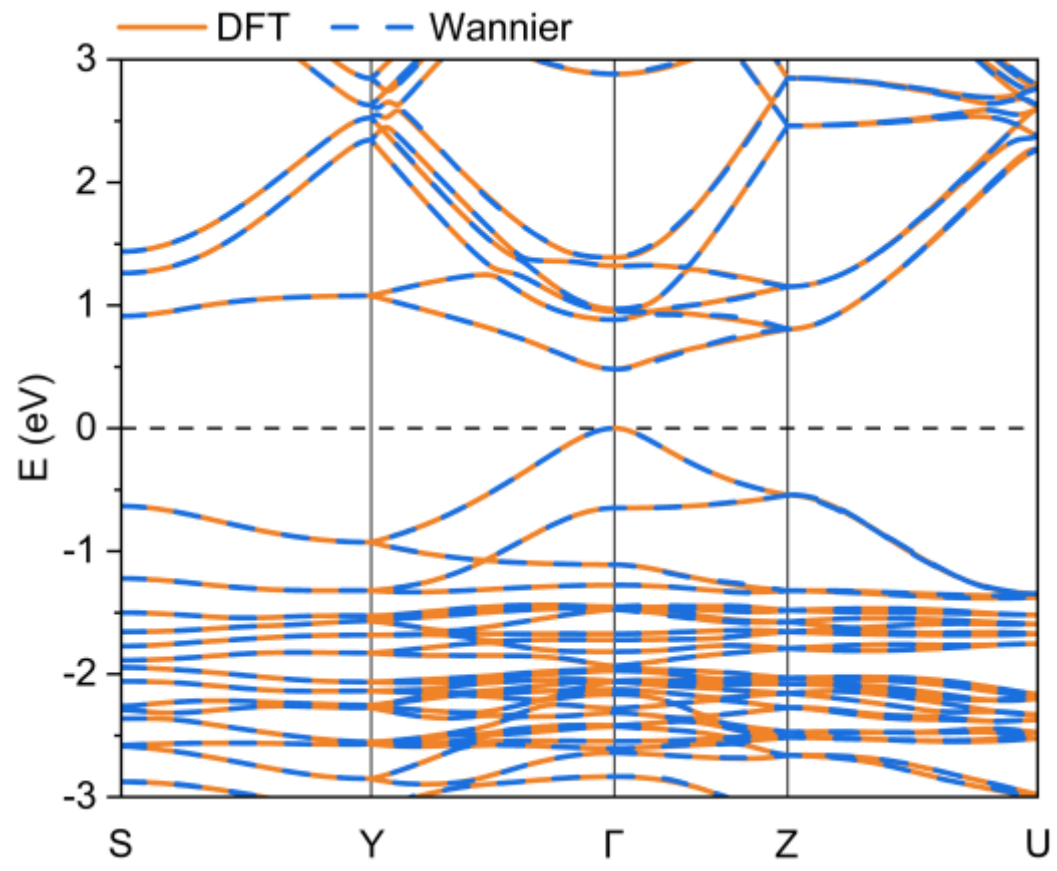

***Figure S2. Wannier-interpolated and DFT band structures of $CsSnI_3$.*** *Bands along S–Y–Γ–Z–U from DFT with spin–orbit coupling (orange solid lines) and from the Wannier interpolation (blue dashed lines). The valence band maximum is set to zero.*

## Convergence of the spin Hall conductivity with the interpolation mesh and the smearing width

Figure S3 shows $\sigma^z_{xy}$ of $CsSnI_3$ as a function of the Fermi level for interpolation meshes from 6×6×4 to 48×48×32 at $\eta$ = 0.05 eV (Figure S3a), and for smearing widths $\eta$ from 1 to 0.05 eV on the 24×24×16 mesh (Figure S3b). The interpolation mesh was refined adaptively by a factor of five at *k*-points where $|\Omega|$ exceeds 100 Å². The curves for the eight meshes coincide throughout the valence band and differ only in the conduction bands 2–3 eV above the gap, so the 24×24×16 mesh used for all calculations is converged. With decreasing $\eta$ the maximum of $\sigma^z_{xy}$, 1.9 eV below the valence band maximum, rises from 214 ($\hbar/e$) S/cm at $\eta$ = 1 eV to 548 ($\hbar/e$) S/cm at $\eta$ = 0.05 eV, and changes by ~5% between $\eta$ = 0.1 and 0.05 eV. Between the gap and ~1 eV below the valence band maximum the curves for the seven widths coincide on the scale of the figure. The origin of this dependence on $\eta$ is discussed in the main text (Eq (2)).

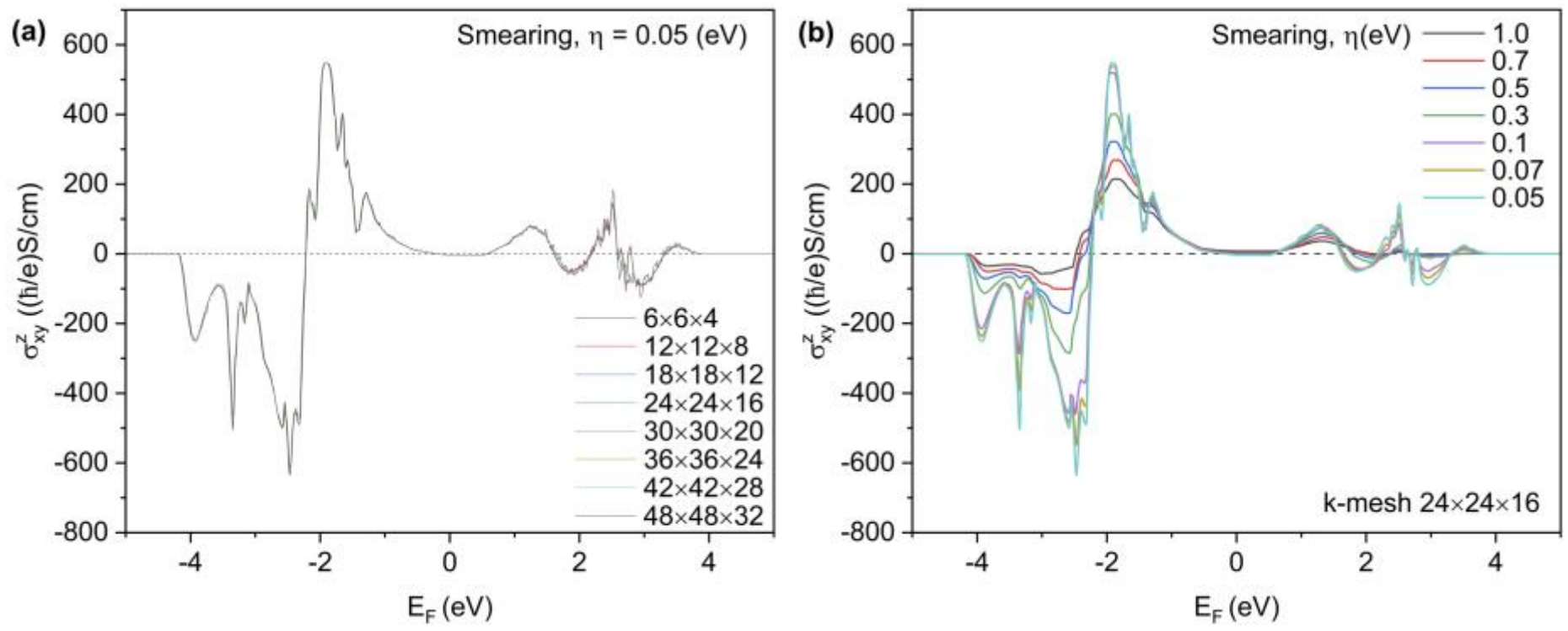

***Figure S3. Convergence of the spin Hall conductivity of $CsSnI_3$ with the interpolation mesh and the smearing width.*** *$\sigma^z_{xy}$ as a function of the Fermi level (a) for interpolation meshes from 6×6×4 to 48×48×32 at $\eta$ = 0.05 eV and (b) for smearing widths $\eta$ from 1 to 0.05 eV on the 24×24×16 mesh. The valence band maximum is set to zero.*

## Components of the spin Hall conductivity

The three calculated components $\sigma^z_{xy}$, $\sigma^x_{yz}$ and $\sigma^y_{zx}$ are independent in the orthorhombic *Pnma* structure (Laue class *mmm*, Methods) and would be equal only in the cubic perovskite structure. Their maximum values differ by 8–26% (Table S2), with $\sigma^z_{xy}$ the smallest in every compound. Their dependence on the Fermi level is similar (Figure S4b and c, compared with Figure 2), as expected for a structure that is a small distortion of the cubic perovskite. For a film with its normal along *z*, the component that drives a spin–orbit torque is $\sigma^y_{zx}$ for a current along *x*, which is calculated here, or $\sigma^x_{zy}$ for a current along *y*, which is not. The Brillouin zone of the *Pnma* cell, with the high-symmetry points used in Figures 1, 3 and S2, is shown in Figure S4a.

***Table S2. Maximum values of the three calculated components of the intrinsic spin Hall conductivity of the four $CsBX_3$ compounds.*** *The maximum of each component with respect to the Fermi level, in (ħ/e) S/cm.*

| Component | $CsPbBr_3$ | $CsPbI_3$ | $CsSnBr_3$ | $CsSnI_3$ |
|---|---|---|---|---|
| $\sigma^z_{xy}$ | 370 | 546 | 318 | 548 |
| $\sigma^x_{yz}$ | 417 | 687 | 344 | 593 |
| $\sigma^y_{zx}$ | 443 | 578 | 385 | 587 |

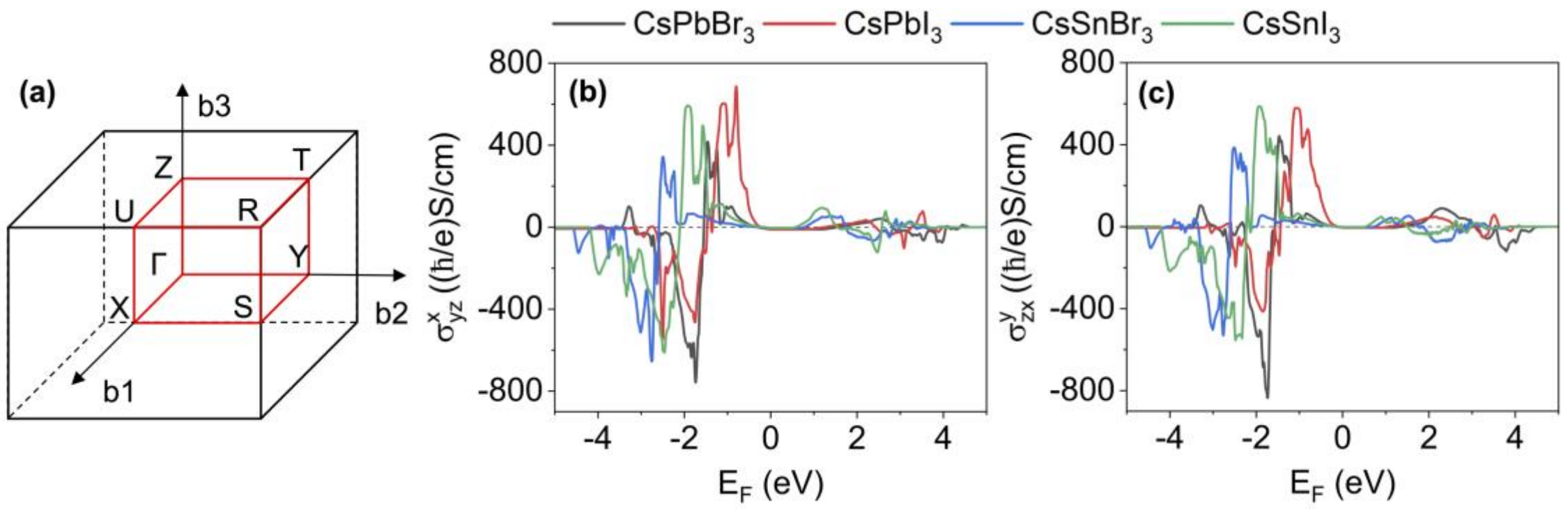

***Figure S4. Brillouin zone and the two other components of the spin Hall conductivity of the four $CsBX_3$ compounds.*** *(a) Brillouin zone of the orthorhombic Pnma cell with the reciprocal lattice vectors* $\mathbf{b}_1$, $\mathbf{b}_2$, $\mathbf{b}_3$ *and the high-symmetry points. (b)* $\sigma^x_{yz}$ *and (c)* $\sigma^y_{zx}$ *as a function of the Fermi level. The valence band maximum is set to zero.*

## Atomic spin–orbit constants

***Table S3. Spectroscopic origin of the atomic spin–orbit constants ζ in Table 1.*** $\zeta = 2\Delta/3$, *where Δ is the fine-structure interval of the ground term of the atom or ion listed, from the NIST Atomic Spectra Database.*[18] *For a single p electron, or a single p hole in the* $p^5$ *configuration of the halogens, Δ is the* $^2P_{1/2}$–$^2P_{3/2}$ *interval. For the* $p^2$ *configuration of* $As^+$ *it is the* $^3P_0$–$^3P_2$ *interval, which is also 3ζ/2 in the Landé limit.*

| Orbital | Species and interval | $\Delta$ (eV) | $\zeta$ (eV) |
|---|---|---|---|
| Pb 6*p* | $Pb^+$, $^2P_{1/2}$–$^2P_{3/2}$ | 1.746 | 1.16 |
| Sn 5*p* | $Sn^+$, $^2P_{1/2}$–$^2P_{3/2}$ | 0.527 | 0.35 |
| I 5*p* | I, $^2P_{3/2}$–$^2P_{1/2}$ | 0.943 | 0.63 |

| Orbital | Species and interval | $\Delta$ (eV) | $\zeta$ (eV) |
|---|---|---|---|
| Br $4p$ | Br, $^2P_{3/2}$–$^2P_{1/2}$ | 0.457 | 0.30 |
| As $4p$ | $As^+$, $^3P_0$–$^3P_2$ | 0.315 | 0.21 |
| Ga $4p$ | Ga, $^2P_{1/2}$–$^2P_{3/2}$ | 0.102 | 0.07 |
| Si $3p$ | $Si^+$, $^2P_{1/2}$–$^2P_{3/2}$ | 0.036 | 0.024 |

## Hole density

The hole density at a Fermi level $E_F$ below the valence band maximum was obtained by integrating the density of states from $E_F$ to the valence band maximum and dividing by the cell volume (Table S1). Table S4 gives the density, and the corresponding number of holes per formula unit, at fixed positions of the Fermi level. Table 2 of the main text gives the inverse, the Fermi level at fixed density.

***Table S4. Hole density of the four $CsBX_3$ compounds as a function of the Fermi level.*** *$E_F$ is measured from the valence band maximum. The hole density p is in $cm^{-3}$. The last column is the number of holes per formula unit. The last row for each compound is at the maximum of $\sigma^z_{xy}$.*

| Compound | $E_F$ (eV) | $p$ ($cm^{-3}$) | Holes per formula unit |
|---|---|---|---|
| $CsPbBr_3$ | −0.1 | $3.6\times10^{19}$ | 0.0075 |
| | −0.3 | $2.2\times10^{20}$ | 0.047 |
| | −0.5 | $8.7\times10^{20}$ | 0.18 |
| | −1.0 | $7.6\times10^{21}$ | 1.59 |
| | −1.43 | $2.9\times10^{22}$ | 6.03 |
| $CsPbI_3$ | −0.1 | $4.4\times10^{19}$ | 0.011 |
| | −0.3 | $5.3\times10^{20}$ | 0.13 |
| | −0.5 | $3.3\times10^{21}$ | 0.82 |
| | −1.0 | $2.3\times10^{22}$ | 5.80 |
| | −1.05 | $2.4\times10^{22}$ | 6.00 |
| $CsSnBr_3$ | −0.1 | $7.3\times10^{18}$ | 0.0015 |
| | −0.3 | $6.5\times10^{19}$ | 0.013 |
| | −0.5 | $1.9\times10^{20}$ | 0.039 |
| | −1.0 | $1.6\times10^{21}$ | 0.32 |
| | −2.58 | $3.6\times10^{22}$ | 7.32 |
| $CsSnI_3$ | −0.1 | $6.6\times10^{18}$ | 0.0016 |
| | −0.3 | $6.7\times10^{19}$ | 0.016 |
| | −0.5 | $2.5\times10^{20}$ | 0.061 |
| | −1.0 | $2.7\times10^{21}$ | 0.66 |
| | −1.94 | $2.6\times10^{22}$ | 6.27 |

## Spin Hall conductivity of $CsSnI_3$ under biaxial strain

Figure S5 shows $\sigma^z_{xy}$ of $CsSnI_3$ as a function of the Fermi level at every strain in the three planes. Table S6 lists the in-gap value, the maximum and its position, and the values at 0.5, 0.74 and 1.0 eV below the valence band maximum. In the unstrained compound these three energies correspond to hole densities of $2.5\times10^{20}$, $10^{21}$ and $2.7\times10^{21}$ $cm^{-3}$ (Table S4, Table 2). The Sn–I–Sn angles at ±5% in the three planes are listed in Table S5.

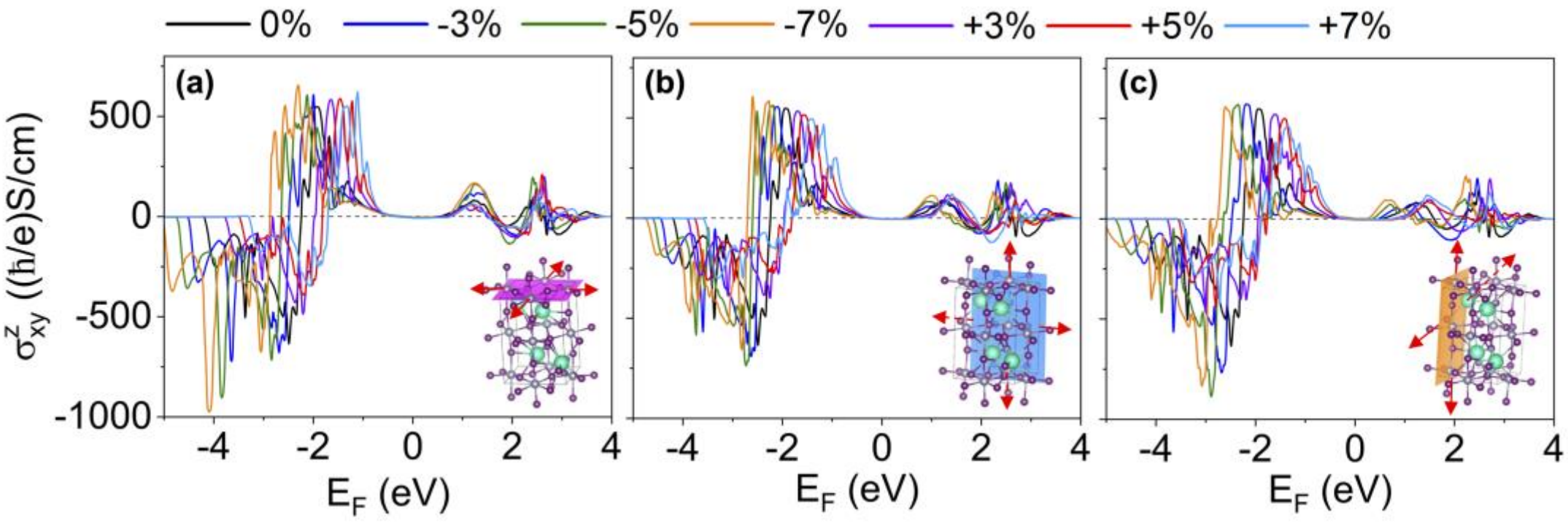

***Figure S5. Spin Hall conductivity of $CsSnI_3$ under biaxial strain in three planes.*** *$\sigma^z{}_{xy}$ as a function of the Fermi level at 0, ±3, ±5 and ±7% biaxial strain in the (a) (001), (b) (010) and (c) (100) planes. The insets show the unit cell with the strained plane shaded. The valence band maximum is set to zero.*

***Table S5. Sn–I–Sn bond angle of $CsSnI_3$ under biaxial strain.*** *The angle between neighbouring $SnI_6$ octahedra at 0 and ±5% strain in the three planes. An angle of 180° would be the untilted cubic arrangement. The angle is the same for every Sn–I–Sn linkage in the cell.*

| Plane | Strain (%) | Sn–I–Sn angle (°) |
|---|---|---|
| – | 0 | 161.5 |
| (001) | −5 | 163.6 |
| (001) | +5 | 159.8 |
| (010) | −5 | 152.8 |
| (010) | +5 | 168.3 |
| (100) | −5 | 152.6 |
| (100) | +5 | 165.5 |

In the (001) plane the Sn–I–Sn angle straightens under compressive strain and bends under tensile strain, whereas in the (010) and (100) planes it bends under compressive strain and straightens under tensile strain (Table S5).

***Table S6. Spin Hall conductivity of $CsSnI_3$ under biaxial strain in the three planes.*** *$\sigma^z{}_{xy}$ in (ħ/e) S/cm at each strain: with the Fermi level in the gap, at its maximum, and at 0.5, 0.74 and 1.0 eV below the valence band maximum. $E_{max}$ is the position of the maximum below the valence band maximum.*

*(001) plane*

| | −7% | −5% | −3% | 0 | +3% | +5% | +7% |
|---|---|---|---|---|---|---|---|
| In the gap | −3.7 | −6.8 | −3.7 | −3.8 | −5.3 | −4.8 | −5.3 |
| Maximum | 655 | 606 | 608 | 548 | 584 | 588 | 622 |
| $E_{max}$ (eV) | −2.30 | −2.12 | −2.00 | −1.94 | −1.64 | −1.46 | −1.11 |
| $\sigma$ at −0.5 eV | 8.5 | 5.9 | 8.5 | 8.9 | 9.0 | 10.5 | 9.0 |
| $\sigma$ at −0.74 eV | 23.8 | 25.8 | 29.9 | 25.8 | 31.9 | 48.2 | 56.9 |
| $\sigma$ at −1.0 eV | 50.8 | 50.3 | 53.8 | 57.0 | 108 | 213 | 223 |

*(010) plane*

| | −7% | −5% | −3% | 0 | +3% | +5% | +7% |
|---|---|---|---|---|---|---|---|
| In the gap | −3.7 | −4.8 | −3.7 | −3.8 | −3.7 | −4.8 | −3.2 |
| Maximum | 607 | 563 | 556 | 548 | 534 | 515 | 495 |
| $E_{max}$ (eV) | −2.60 | −2.19 | −2.09 | −1.94 | −1.69 | −1.56 | −1.43 |

| | −7% | −5% | −3% | 0 | +3% | +5% | +7% |
|---|---|---|---|---|---|---|---|
| $\sigma$ at −0.5 eV | 7.5 | 5.9 | 6.5 | 8.9 | 15.6 | 26.8 | 29.9 |
| $\sigma$ at −0.74 eV | 12.1 | 13.6 | 16.6 | 25.8 | 43.6 | 61.5 | 98.6 |
| $\sigma$ at −1.0 eV | 34.0 | 37.0 | 38.5 | 57.0 | 127 | 228 | 226 |

*(100) plane*

| | −7% | −5% | −3% | 0 | +3% | +5% | +7% |
|---|---|---|---|---|---|---|---|
| In the gap | 0.9 | −2.2 | −4.2 | −3.8 | −4.8 | −4.8 | −3.2 |
| Maximum | 558 | 569 | 572 | 548 | 520 | 499 | 460 |
| $E_{max}$ (eV) | −2.60 | −2.34 | −2.19 | −1.94 | −1.62 | −1.41 | −1.34 |
| $\sigma$ at −0.5 eV | 11.5 | 8.5 | 7.0 | 8.9 | 13.1 | 30.4 | 43.6 |
| $\sigma$ at −0.74 eV | 14.1 | 15.1 | 15.1 | 25.8 | 49.2 | 109 | 170 |
| $\sigma$ at −1.0 eV | 26.3 | 29.9 | 36.5 | 57.0 | 132 | 168 | 154 |

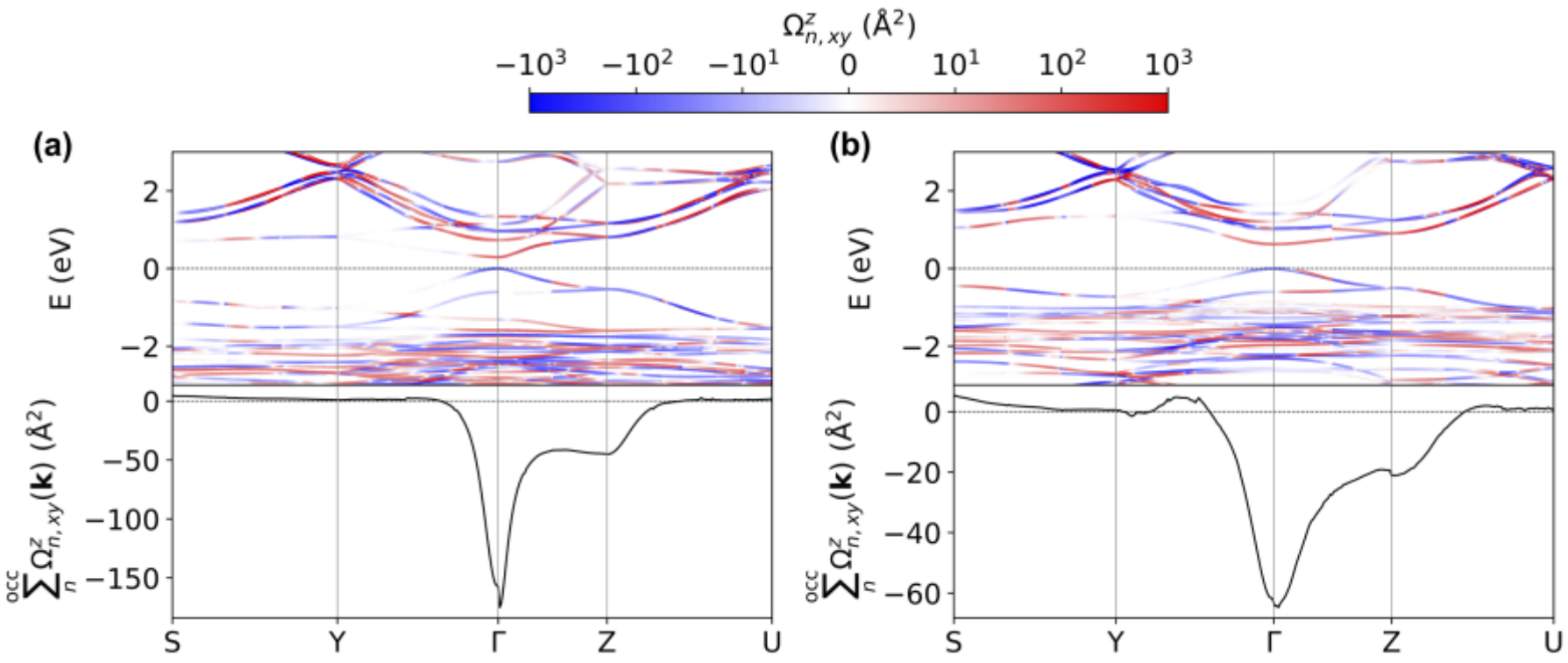

***Figure S6. Spin Berry curvature of $CsSnI_3$ under biaxial strain.*** *Band structure of $CsSnI_3$ along S–Y–Γ–Z–U with spin–orbit coupling at (a) −5% and (b) +5% biaxial strain in the (001) plane, each state coloured by its spin Berry curvature $\Omega^z_{n,xy}$(**k**) on a logarithmic scale for each sign (upper panels), and $\Omega^z_{xy}$(**k**) summed over the valence bands (lower panels). The valence band maximum is set to zero.*